\documentclass[aps,twocolumn,superscriptaddress]{revtex4}
\usepackage{graphicx}
\usepackage{tabularx}
\usepackage{makecell}
\usepackage{xcolor}
\usepackage{ulem}
\usepackage{multirow}
\usepackage{booktabs}
\begin{document}

\title{Multiband superconductivity and a deep gap minimum from the specific heat in KCa$_2$(Fe$_{1-x}$Ni$_x$)$_4$As$_4$F$_2$ ($x$ = 0, 0.05, 0.13) }

\author{Yiwen Li}
\affiliation{National Laboratory of Solid State Microstructures and Department of Physics, Collaborative Innovation Center of Advanced Microstructures, Nanjing University, Nanjing 210093, China} 

\author{Zhengyan Zhu}
\affiliation{National Laboratory of Solid State Microstructures and Department of Physics, Collaborative Innovation Center of Advanced Microstructures, Nanjing University, Nanjing 210093, China} 

\author{Yongze Ye}
\affiliation{National Laboratory of Solid State Microstructures and Department of Physics, Collaborative Innovation Center of Advanced Microstructures, Nanjing University, Nanjing 210093, China} 

\author{Wenshan Hong}
\affiliation{Beijing National Laboratory for Condensed Matter Physics, Institute of Physics, Chinese Academy of Sciences, Beijing 100190, China}
\affiliation{School of Physical Sciences, University of Chinese Academy of Sciences, Beijing 100190, China}
\affiliation{International Center for Quantum Materials, School of Physics, Peking University, Beijing 100871, China}

\author{Yang Li}
\affiliation{Beijing National Laboratory for Condensed Matter Physics, Institute of Physics, Chinese Academy of Sciences, Beijing 100190, China}
\affiliation{School of Physical Sciences, University of Chinese Academy of Sciences, Beijing 100190, China}

\author{Shiliang Li}
\affiliation{Beijing National Laboratory for Condensed Matter Physics, Institute of Physics, Chinese Academy of Sciences, Beijing 100190, China}
\affiliation{School of Physical Sciences, University of Chinese Academy of Sciences, Beijing 100190, China}

\author{Huiqian Luo}\thanks{hqluo@iphy.ac.cn} 
\affiliation{Beijing National Laboratory for Condensed Matter Physics, Institute of Physics, Chinese Academy of Sciences, Beijing 100190, China}

\author{Hai-Hu Wen}\thanks{hhwen@nju.edu.cn}
\affiliation{National Laboratory of Solid State Microstructures and Department of Physics, Collaborative Innovation Center of Advanced Microstructures, Nanjing University, Nanjing 210093, China}

\date{\today}

\begin{abstract}
Specific heat can explore low-energy quasiparticle excitations of superconductors, so it is a powerful tool for bulk measurement on the superconducting gap structure and pairing symmetry. Here, we report an in-depth investigation on the specific heat of the multiband superconductors KCa$_2$(Fe$_{1-x}$Ni$_x$)$_4$As$_4$F$_2$ ($x$ = 0, 0.05, 0.13) single crystals and the overdoped non-superconducting one with $x$ = 0.17. Clear specific heat anomalies can be observed at the superconducting transition temperature of 33.6 K and 28.8 K for the samples with $x$ = 0 and $x$ = 0.05, respectively. For the two samples, the magnetic field induced specific heat coefficient $\Delta\gamma(H)$ in the low-temperature limit increases rapidly below 2 T, then it rises slowly above 2 T. Using the non-superconducting sample with $x$ = 0.17 as a reference, the specific heat of phonon background for various superconducting samples can be obtained and subtracted, which allows us to extract the electronic specific heat of the superconducting samples. Through comparative analyses, it is found that the energy gap structure including two $s$-wave gaps and an extended $s$-wave gap with large anisotropy can reasonably describe the electronic specific heat data. According to these results, we suggest that at least one anisotropic superconducting gap with a deep gap minimum should exist in this multiband system. With the doping of Ni, the superconducting transition temperature of the sample decreases along with the decrease of the large $s$-wave gap, but the extended $s$-wave gap increases due to the enlarged electron pockets via adding more electrons. Despite these changes, the general properties of the gap structure remain unchanged versus doping Ni. In addition, the calculation of condensation energy of the parent and doped samples shows the rough consistency with the correlation of $U_0 \propto {T_c}^n$ with $n$ = 3-4, which is beyond the understanding of the BCS theory.

\end{abstract}

\maketitle

\section{Introduction}
	Since the discovery of superconductivity in LaFeAsO$_{1-x}$F$_x$ \cite{RN1} in 2008, the study on iron-based superconductors (IBSs) has attracted great attention due to their remarkable properties, such as high superconducting transition temperature ($T_c$) and high upper critical field ($H_{c2}$). As these properties are beneficial for future applications, it is highly desirable for people to search for new IBSs. Up to now, they have found many different types of IBSs, such as 1111-type (LaFeAsO$_{1-x}$F$_x$) \cite{RN1}, 122-type (Ba$_{1-x}$K$_x$Fe$_2$As$_2$) \cite{RN2}, 111-type (LiFeAs) \cite{RN3} and 11-type ($\alpha$-FeSe) \cite{RN4}. It is commonly recognized that IBSs have multi-orbital characteristics, resulting in complicated electronic energy bands and Fermi surface structures, which cannot be described within the single-band model. In 2016, the KCa$_2$Fe$_4$As$_4$F$_2$ superconductors were discovered \cite{RN5}, which initiates the new family named 12442-type with the general formula $A$Ca$_2$Fe$_4$As$_4$F$_2$ ($A$ = K, Rb and Cs) \cite{RN5,RN6}. They are abbreviated as K12442, Rb12442, and Cs12442, respectively. This system possesses a unique and hybrid crystal structure that combines alternative 122-type and 1111-type structures. It comprises double conducting Fe$_2$As$_2$ layers between insulating Ca$_2$F$_2$ layers, making this system the only IBSs with this unique structure and resulting in strong quasi-two-dimensional characteristics. To study the structural and electronic properties of the system, Wang. et. al have done the first principles calculations \cite{RN7}. The calculations present ten bands crossing the Fermi level, with six hole-like pockets at the center of the Brillouin zone and four electron-like pockets at the corner of the zone. The result reveals that the 12442 system possesses more complicated Fermi surface structures than most IBSs and exhibits multigap properties.\par 
	Concerning the gap structures in IBSs, there is a consensus that, in general, IBSs have nodeless $s$-wave pairing symmetry, except for a few systems such as LaFePO, BaFe$_2$(As$_{0.7}$P$_{0.3}$)$_2$ and KFe$_2$As$_2$, which may have nodes \cite{RN8,RN9,RN10}. However, the gap structure in the 12442 system, although studied by many different methods, remains controversial. Thermal conductivity measurements \cite{RN11} and optical spectroscopy \cite{RN12} suggest the multigap nodeless superconductivity in Cs12442 single crystals. Scanning tunneling microscopy/spectroscopy (STM/STS) \cite{RN13} indicates full-gap features, and angle-resolved photoemission (ARPES) \cite{RN14} experiment shows that the superconducting gaps are almost isotropic in K12442 single crystals. On the contrary, $\mu$SR experiments \cite{RN15,RN16} on polycrystalline samples of K12442 and Cs12442 have revealed the existence of line nodes in the superconducting gaps. Later, directional point-contact Andreev-reflection spectroscopy \cite{RN17} and specific heat measurements \cite{RN18} have also provided evidence for the possible presence of nodal gaps in single crystals Rb12442 and K12442. These studies with different conclusions reflect the complexity of the system, and systematic investigations on high-quality samples using different techniques are highly desired.\par
	In this work, electrical transport and thermodynamic studies have been carried out on KCa$_2$(Fe$_{1-x}$Ni$_x$)$_4$As$_4$F$_2$ ($x$ = 0, 0.05, 0.13, 0.17) single crystals. In the 12442 system, Ni acts as an electron dopant. It has been proved by Yi et al. \cite{RN19} that the substitution of Fe with Ni brings electrons into the 12442 system and introduces disorders that contribute to the pair breaking effect at the same time. Specific heat is an important tool to provide bulk evidence for the pairing symmetry and gap structure. Specific heat study of the low-energy excitations for K12442 single crystals has been reported by Wang et al. \cite{RN18}, and proposed a possible $d$-wave gap. However, due to the lack of electronic specific heat data, they cannot fit it with different gap functions, which makes the gap structure unsettled. Here, we report an in-depth analysis on the specific heat of KCa$_2$(Fe$_{1-x}$Ni$_x$)$_4$As$_4$F$_2$ ($x$ = 0, 0.05, 0.13, 0.17) single crystals. By subtracting the phonon contribution, we obtain the electronic specific heat data and then fit them with two $s$-wave gaps plus an extended $s$-wave gap with large anisotropy. The fitted results, combined with the behavior of the magnetic field induced specific heat coefficient $\Delta\gamma(H)$, which increases quickly at low magnetic field region, show that at least one anisotropic superconducting gap with deep gap minima exists in this multiband system. Ni doping only changes the size of the gaps derived from the fitting, but does not change its general gap structure. Moreover, we calculate the condensation energy of the three superconducting samples, and find the correlation of $U_0 \propto{T_c}^{3.5 \pm 0.5}$, which is found in several iron-based systems as well \cite{RN20}. This phenomenon may be related to the mechanism of unconventional superconductivity. Our results provide new information about the energy gap structure and pairing symmetry of this unconventional superconducting system.

\section{Materials and methods}

Single-crystalline KCa$_2$(Fe$_{1-x}$Ni$_x$)$_4$As$_4$F$_2$ ($x$ = 0, 0.05, 0.13, 0.17) were synthesized by the self-flux method. The details about the growth method can be found in Ref. \cite{RN21}. The samples with $x$ = 0, 0.5, 0.13, 0.17 for the specific heat measurements have masses of 2.12 mg, 2.06 mg, 1.92 mg, and 0.87 mg, respectively. The magnetization was measured with a superconducting quantum interference device (SQUID-VSM, Quantum Design). The resistivity and the specific heat were measured with a physical property measurement system (PPMS, Quantum Design) by standard four-probe method and thermal-relaxation method, respectively. During the specific heat measurement, different magnetic fields up to 9 T were applied along the $c$-axis of the samples. Before doing measurements on each sample, the specific heat of the sample holder and the addenda (Apiezon N Grease) were premeasured and subtracted away, thus we report here only the specific heat from the samples. 

\section{Results and discussion}
\subsection{Resistivity and magnetization characterization}

\begin{figure}[htbp]
	\includegraphics[width=7cm]{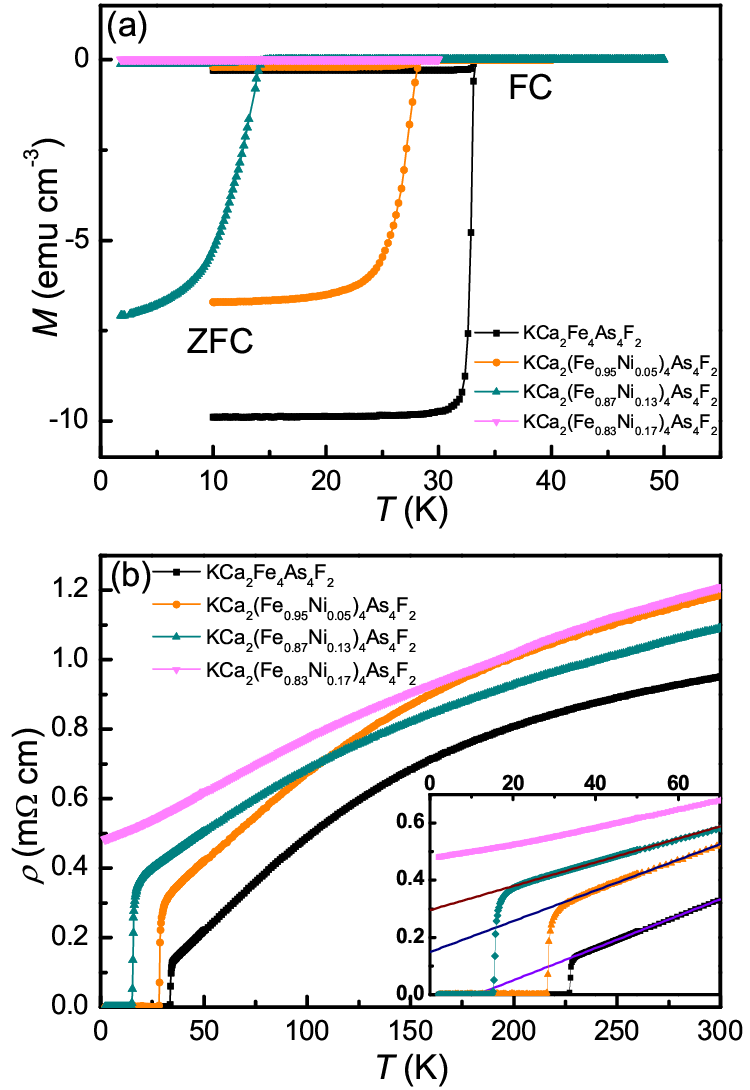}
	\caption{(a) Temperature dependence of the magnetization of KCa$_2$(Fe$_{1-x}$Ni$_x$)$_4$As$_4$F$_2$ ($x$ = 0, 0.05, 0.13, 0.17) single crystals under a magnetic field of 5 Oe in zero-field-cooling (ZFC) and field-cooling (FC) modes. (b) Temperature dependence of the in-plane resistivity of KCa$_2$(Fe$_{1-x}$Ni$_x$)$_4$As$_4$F$_2$ ($x$ = 0, 0.05, 0.13, 0.17) single crystals at zero magnetic field. The inset shows the resistivity data from 2 K to 70 K. The solid lines are the linear fitting results for the superconducting samples.} \label{fig1}
	\end{figure}

Figure~\ref{fig1}(a) shows the magnetization of KCa$_2$(Fe$_{1-x}$Ni$_x$)$_4$As$_4$F$_2$ ($x$ = 0, 0.05, 0.13, 0.17) at 5 Oe for both zero-field cooling (ZFC) and field-cooling (FC) modes. The diamagnetic transitions are detected at 33.3 K, 28.6 K, and 14.5 K for the samples with $x$ = 0, 0.05, 0.13, respectively. After being corrected by the demagnetization effect, the samples roughly have full magnetic shielding volume in the low-temperature limit. Figure~\ref{fig1}(b) presents the in-plane resistivity measurements of the samples KCa$_2$(Fe$_{1-x}$Ni$_x$)$_4$As$_4$F$_2$ ($x$ = 0, 0.05, 0.13, 0.17) at zero magnetic field from 2 K to 300 K. Using the criterion of 50$\%$ of the normal state resistivity, the $T_c$ can be determined as 33.9 K, 28.7 K, and 15.7 K for the three superconducting samples, respectively. The superconducting transitions are quite sharp, demonstrating the good quality of our samples. For the samples with $x$ = 0.17, both resistivity and magnetization measurements do not detect any superconducting behavior above 2 K. It is found that upon Ni doping, $T_c$ decreases gradually until superconductivity is killed.\par
	In order to obtain more information about their electric transport properties, we can study the resistivity in the normal state. Remarkably, the resistivity data for the three superconducting samples exhibit a good linear relationship with temperature at low temperatures, indicating a non-Fermi liquid behavior. To analyze this behavior, we adopt a linear fitting approach to the resistivity data of the three superconducting samples below 70 K, using Equation (1):
     \begin{equation}
\rho_{ab} = \rho_0 + AT \label{equation1}.
     \end{equation}

Here, $\rho_{ab}$ represents the in-plane resistivity, $\rho_0$ is the residual resistivity which accounts for defect and impurity scattering, and $AT$ represents the resistivity contributed by non-Fermi liquid scattering mechanism. In this low-temperature range, the resistivity contribution from electron-phonon scattering, which is typically a power-law function with a high-order exponent of temperature, is negligible and can be ignored. The fitted curves, represented by solid lines, are illustrated in the inset of Fig.~\ref{fig1}(b). The excellent fitting results of all data further support this non-Fermi liquid behavior. However, it is worth noting that, for the parent sample, the resistivity intercept at zero temperature yields a negative value, which is unphysical. This phenomenon has also been observed in Rb12442, Cs12442, and other IBSs \cite{RN19,RN22,BKFA}. It implies that the resistivity curve should display significant positive curvature at low temperatures, as observed in the non-superconducting sample with $x$ = 0.17. This positive curvature at low temperatures can be attributed to the existence of a $T^2$ term contributed by the Fermi liquid scattering mechanism. Considering the multi-orbital feature of the system, it is expected that multiple types of carriers are involved in the electric conduction. While the dominant contribution is presumed to be the $T$-linear term associated with non-Fermi liquid behavior, additional terms, such as a $T^2$ term, are expected to simultaneously participate in electric conduction, to prevent a negative intercept.
Additionally, we observe an enhancement in the residual resistivity $\rho_0$ with doping. This can be attributed to the enhanced impurity scattering effect caused by the substituted Ni atoms, which also explains the reason why $T_c$ decreases with the doping level.

\subsection{Specific heat measurements under different magnetic fields}

\begin{figure}[htbp]
	\includegraphics[width=7cm]{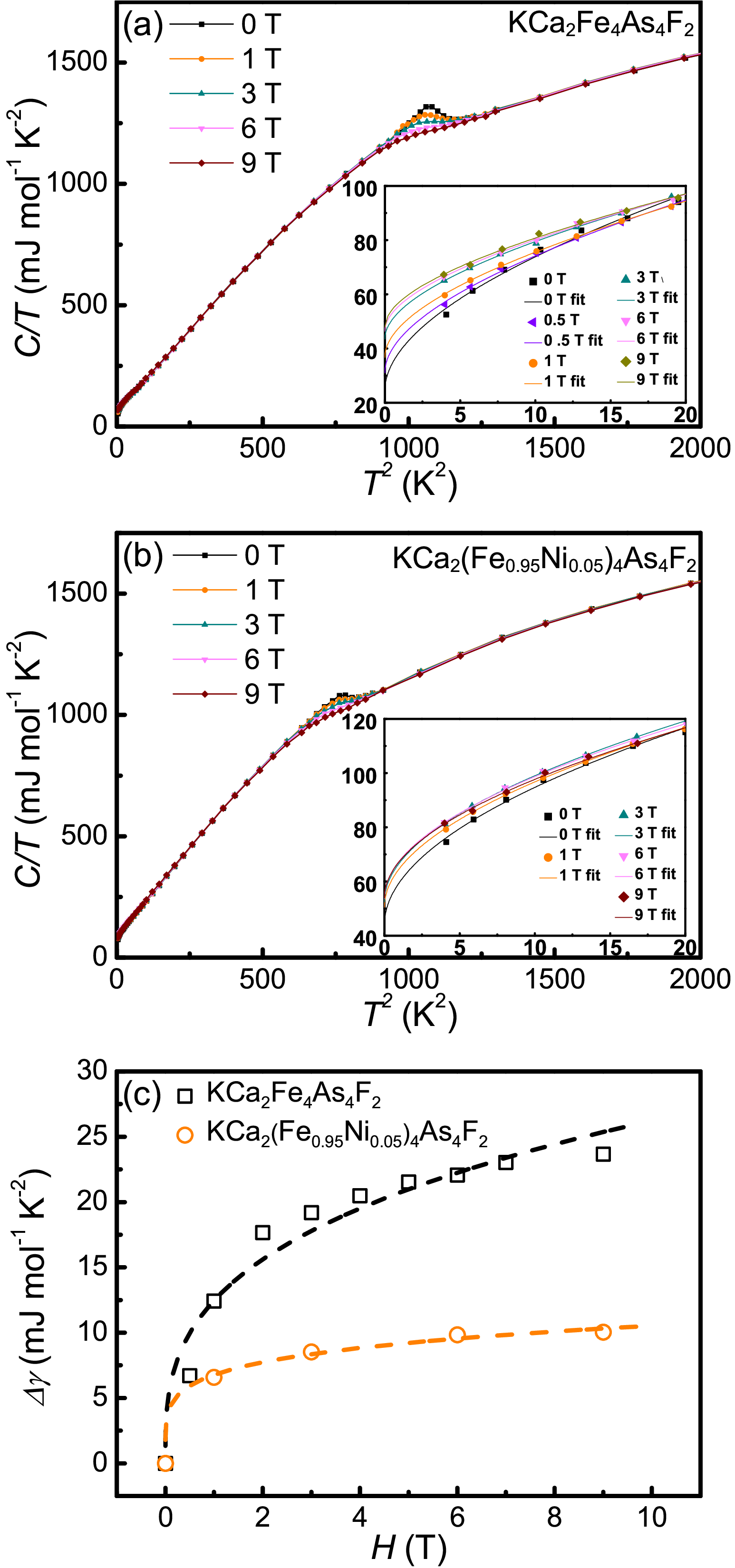}
	\caption{Temperature dependence of specific heat plotted as $C/T$ vs $T^2$ for KCa$_2$(Fe$_{1-x}$Ni$_x$)$_4$As$_4$F$_2$ with (a) $x$ = 0 and (b) $x$ = 0.05 under different magnetic fields. The insets are the low-temperature specific heat of the corresponding samples from 2 K to 4.5 K. Solid lines show the fitted curves (see text). (c) The magnetic field induced specific heat coefficient for KCa$_2$(Fe$_{1-x}$Ni$_x$)$_4$As$_4$F$_2$ with $x$ = 0 and $x$ = 0.05. The dashed lines show the fitted curves (see text).} \label{fig2}
	\end{figure}
	We have measured the specific heat of KCa$_2$(Fe$_{1-x}$Ni$_x$)$_4$As$_4$F$_2$ ($x$ = 0, 0.05, 0.13, 0.17) and the results are presented in Fig.~\ref{fig3}(a). In Figs.~\ref{fig2}(a) and 2(b), we plot the $C/T-T^2$ relationship for the samples with $x$ = 0 and $x$ = 0.05 from 2 K to 45 K under different magnetic fields. The specific heat jumps appear at 33.6 K and 28.8 K for these two samples, respectively. When applying a magnetic field, the jump becomes less obvious. However, for the sample with $x$ = 0.13, the jump is too small to be detected. Nonetheless, when subtracting the huge phonon background from the total specific heat, the jump evolves into a hump but can still be observed, as shown in Fig.~\ref{fig3}(b). To further study the low-energy excitations, we plot the low-temperature behavior from 2 K to 4.5 K in the insets of Figs.~\ref{fig2}(a) and 2(b). It can be seen that there are slight deviations from the linear behavior for the relation between $C/T$ and $T^2$, so we cannot simply describe the curves with $C(T,H) = \gamma_0(H)T+\beta T^3$ where the first term is electronic contribution owing to magnetic field induced excitations and the second term is phonon contribution due to lattice vibrations. In this case, we add a field-dependent term $\alpha(H)T^2$ to the electronic contribution, so that the formula becomes $C(T,H) = \gamma_0(H)T + \alpha(H)T^2 + \beta T^3$. This $\alpha(H)T^2$ term usually means the existence of line nodes \cite{RN23}. However, from another point of view, considering the multiband feature of the 12442 system, it may be generated from another relatively small energy gap as well. For the samples with $x$ = 0 and $x$ = 0.05, the fitted curves with the above formula under different fields are represented by solid lines in the insets of Figs.~\ref{fig2}(a) and 2(b). For the sample with $x$ = 0.13, there is not much difference between the specific heat data of 0 T and 9 T, so we are not going to discuss it in the following part. From the fitted parameters, we have the zero-temperature specific heat coefficients $\gamma_0(H)$ under different magnetic fields. Fig.~\ref{fig2}(c) exhibits the field-induced term $\Delta\gamma(H) = \gamma_0(H) - \gamma_0(0)$ of the two samples, which reflects the information about the superconducting gap. It can be seen that $\Delta\gamma(H)$ increases rapidly with the magnetic field at first below 2 T and then the ramping rate slows down above 2 T. Actually, the residual specific heat in the superconducting state $\gamma_0(H)$ reflects the field-dependent density of states (DOS) near the Fermi surface. As the magnetic field increases, superconductivity is gradually suppressed and more quasiparticles are excited, leading to an increase in the $\gamma_0$ value. For different gap structures, $\gamma_0$ values will increase with the magnetic field in different ways. Therefore, studying the behavior of $\Delta\gamma(H)$ is very helpful to study its gap structure. Usually, for an isotropic $s$-wave gap, quasiparticle excitations arise only from vortex cores, and the density of vortices is proportional to the magnetic field, so $\Delta\gamma(H)$ versus $H$ is a linear relation. For a $d$-wave gap, $\Delta\gamma(H)$ follows a square root tendency with magnetic field because of the Doppler shift effect where the superconducting condensate outside the vortex cores will contribute to the DOS as well \cite{RN24}. The dashed lines in Fig.~\ref{fig2}(c) are the power law fits of the two samples using $\Delta\gamma(H) = aH^b$, which yields a power exponent $b$ of 0.31 and 0.19, respectively ($a$ and $b$ are the fitted parameters). The fittings do not agree with a $d$-wave gap or an $s$-wave gap, but the results strongly suggest the existence of some deep minima in the gap function. In combination with the $\alpha(H)T^2$ term, we are more convinced that the gap should have highly anisotropic characteristics. 

\subsection{Extraction of the phonon contribution from specific heat}

\begin{figure}[htbp]
	\includegraphics[width=7cm]{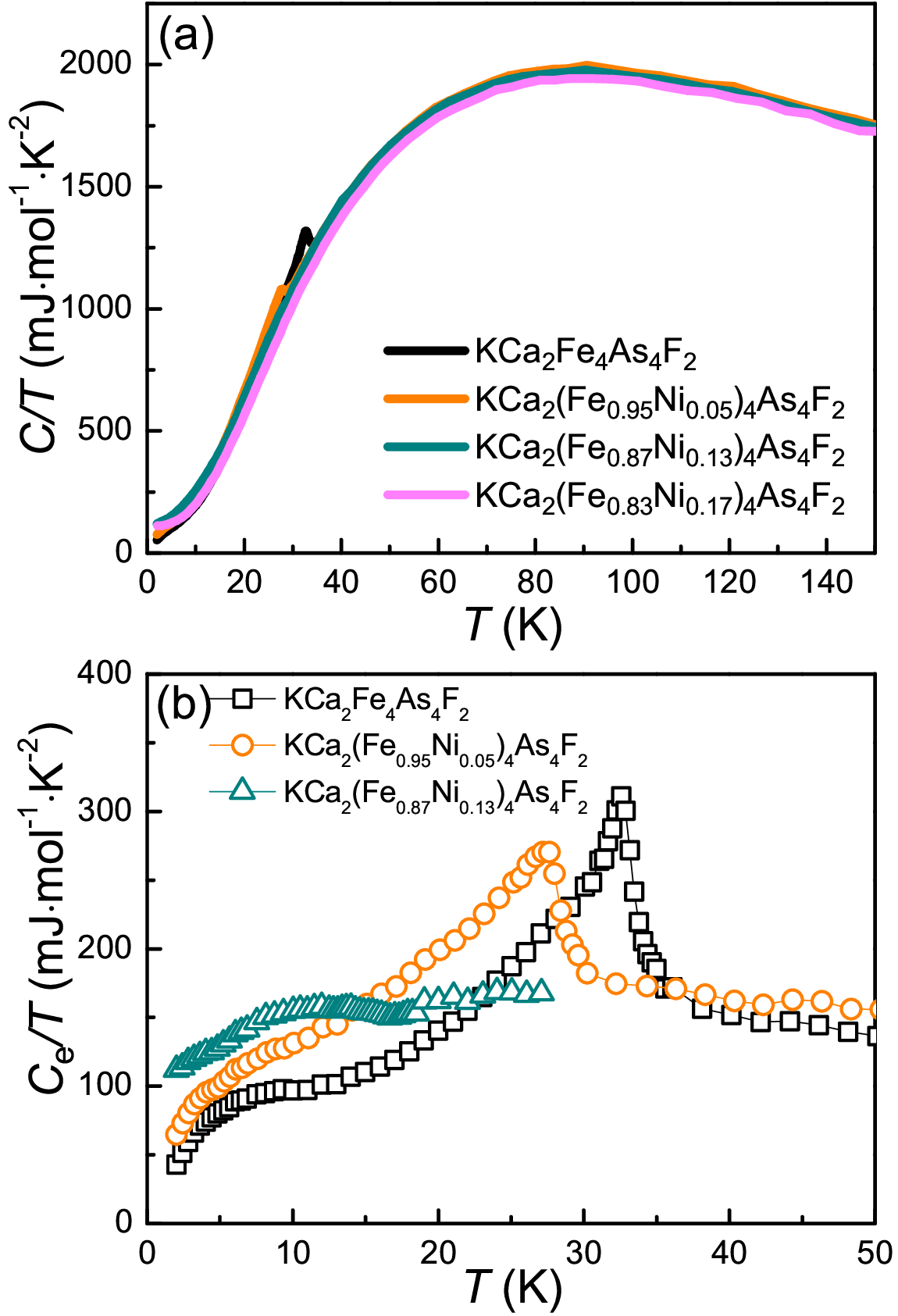}
	\caption{(a) Temperature dependence of specific heat for KCa$_2$(Fe$_{1-x}$Ni$_x$)$_4$As$_4$F$_2$ ($x$ = 0, 0.5, 0.13, 0.17) from 2 K to 150 K. (b) Temperature dependence of the electronic specific heat for KCa$_2$(Fe$_{1-x}$Ni$_x$)$_4$As$_4$F$_2$ with $x$ = 0, 0.5, 0.13.} \label{fig3}
	\end{figure}

	In a superconductor, specific heat usually consists of phonon and electronic contributions. The phonon contribution of specific heat is caused by lattice vibrations and the electronic part is related to the electronic near the Fermi surface. Electronic specific heat behaves differently in the superconducting state compared to the normal state. Analyzing the behavior of the electronic specific heat in the superconducting state is necessary to study the electron pairing mechanism. Therefore, it is very important to extract electronic term from the total specific heat. Usually, we apply a high magnetic field to suppress superconductivity or follow the simplified Debye model ($C = \beta T^3$) to obtain the normal state specific heat in low-temperature region. However, for the high-$T_c$ superconductors like the 12442 system, the upper critical field is too high to reach and the simplified Debye model is inapplicable for the fitting in high-temperature range. In this case, it is a common practice to use non-superconducting samples that share similar crystal structures and element compositions as references to extract the phonon background of the system. These reference samples are assumed to display phonon background behaviors similar to those of superconducting materials. In our study, we have grown overdoped samples with $x$ = 0.17 that do not exhibit superconductivity above 2 K, and we assemble several pieces with $x$ = 0.17 for the specific heat measurements in order to enhance the total signal. The specific heat from this assembled overdoped sample can serve as the reference for determining the phonon background. We have measured the specific heat of the samples with varying Ni doping levels ($x$ = 0, 0.05, 0.13, 0.17), as depicted in Fig.~\ref{fig3}(a). Remarkably, these samples exhibit very similar specific heat behavior above $T_c$. This consistency allows us to effectively utilize the non-superconducting sample as a phonon reference. However, there may still be minor differences in the phonon contribution of specific heat due to factors such as mass error and atom substitution. To account for these subtle differences and ensure the accuracy of our analysis, we have implemented a method suggested in Ref. \cite{ZB} for adjusting the reference sample data to match the phonon background of the different superconducting samples.\par
	There is an assumption that the entropy of samples with varying levels of doping conforms to a universal form: $S(lattice) = \epsilon\Phi(T/\theta)$ with $\theta$ and $\epsilon$ representing characteristic parameters specific to each sample, and $\Phi$ being a function common to all samples. According to the thermodynamic formula, the phonon specific heat can be expressed as $C_{ph}(T) = T\partial{S(lattice)}/\partial{T}  = (\epsilon/\theta)T\Phi'(T/\theta)$. Therefore, for different superconducting samples, they are related to the non-superconducting sample by the equation $C_{ph-s}(T) = a\cdot C_{ph-n}(b \cdot T)$. Here, $C_{ph-s}(T)$ and $C_{ph-n}(T)$ denote the phonon contribution of specific heat from the superconducting and non-superconducting samples, respectively; $a$ and $b$ are fitting parameters. By subtracting a constant, representing the electronic contribution, from the total specific heat coefficient $C_{total}/T$ of the non-superconducting sample, we can obtain its phonon contribution. Then, by fine-tuning parameters $a$ and $b$, while ensuring entropy conservation, we can accurately determine the phonon contributions of the superconducting samples. The results are $a$ and $b$ to be 0.99 and 1 for the sample with $x$ = 0; 0.985 and 0.99 for the sample with $x$ = 0.05; and 0.99 and 0.98 for the sample with $x$ = 0.13.\par
	By subtracting the phonon contributions from the experimental data, we ultimately obtain the electronic specific heat for superconducting samples, as depicted in Fig.~\ref{fig3}(b). In the next part, we will explore the information contained in the electronic specific heat and further study the gap structure of the 12442 system. One peculiar behavior of the electronic specific heat is that there is a clear bending down below about 10 K.\par

\subsection{Electronic specific heat fitted with different gap structures}
\begin{figure}[htbp]
	\includegraphics[width=7cm]{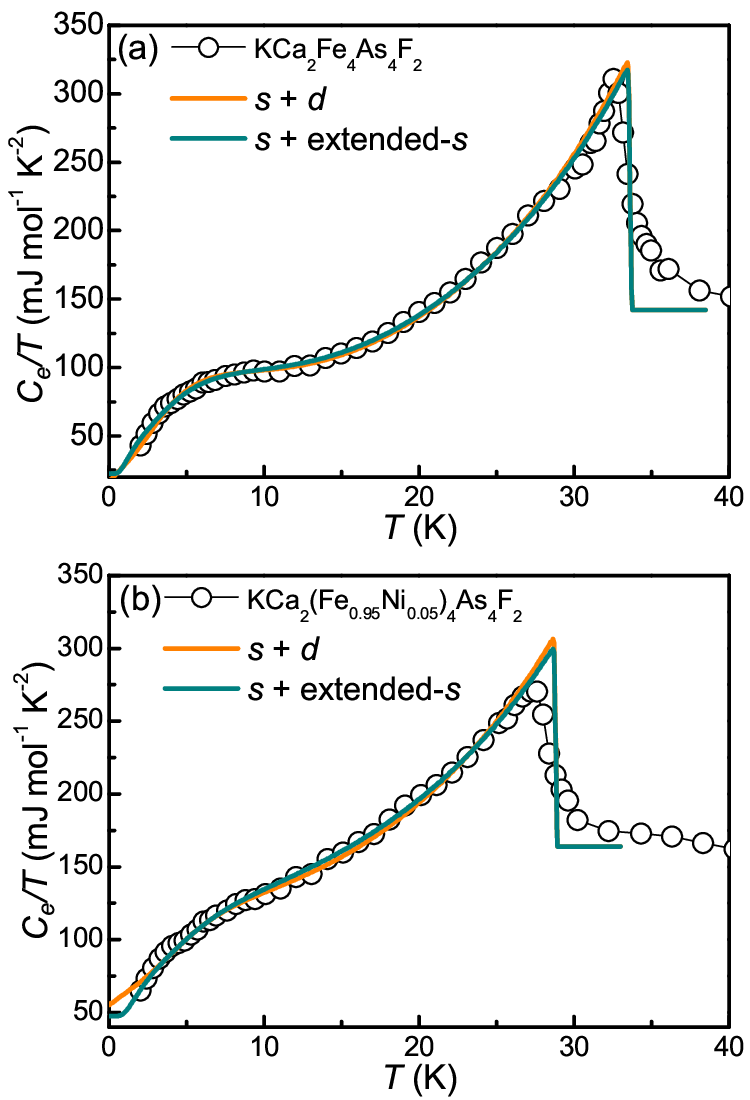}
	\caption{Temperature dependence of the electronic specific heat fitted with a two-gap model for (a) KCa$_2$Fe$_4$As$_4$F$_2$ and (b) KCa$_2$(Fe$_{0.95}$Ni$_{0.05}$)$_4$As$_4$F$_2$. The lines represent the fitted curves with the BCS model.} \label{fig4}
	\end{figure}
	While many researchers have focused on the pairing symmetry of the superconducting gap in the 12442 system, there is still no consensus on whether the gap has nodes or not and whether the nodes are accidental or not. Electronic specific heat contains useful information about superconducting gap structure. By fitting the electronic specific heat data with different gap structures, we will have a deep understanding of the specific composition of the energy gap, which provides another experimental evidence for the conclusion of the gap structure in the 12442 system. The fitting formula is based on the BCS theory, which follows as
     \begin{eqnarray}
\gamma_e' = \frac{4N(E_F)}{k_B T^3}\int_{0}^{+\infty}\int_{0}^{2\pi}\frac{e^{\xi/k_BT}}{(1 + e^{\xi/k_BT})^2}\\ 
(\epsilon^2+\Delta^2(\theta,T) - \frac{T}{2}\frac{d\Delta^2(T,\theta)}{dT})d\theta d\epsilon. \label{equation5}	\nonumber
     \end{eqnarray}
Here, $\gamma_e' = \gamma_e - \gamma_0$, which defines the electronic specific heat coefficient after subtracting the residual component at zero temperature. $\xi = \sqrt{\epsilon^2 + \Delta^2(T,\theta)}$. The kinetic energy $\epsilon = \hbar^2k^2/2m - E_F$. $\Delta(T,\theta)$ is the gap function, and it has different forms based on different gap structures. We list it as (i) a single $s$-wave gap $\Delta(T,\theta) = \Delta_0(T)$, (ii) a single $d$-wave gap $\Delta(T,\theta) = \Delta_0(T)\cos2\theta$, (iii) a single extended $s$-wave gap $\Delta(T,\theta) = \Delta_0(T)(1 + \alpha \cos2\theta)$. Here, $\alpha$ denotes the anisotropy. If $\alpha$ = 1, the gap function will be zero at some points, indicating that there exist accidental nodes. When $\alpha$ is close to 0, it means the anisotropy is very small and the extended $s$-wave gap will degenerate into an isotropic $s$-wave gap.  \par
	We first consider the one-gap model to fit the electronic specific heat data, and find that for both KCa$_2$Fe$_4$As$_4$F$_2$ and KCa$_2$(Fe$_{0.95}$Ni$_{0.05}$)$_4$As$_4$F$_2$, a single gap cannot fit the data well (see Fig.~\ref{fig7}(a) and 7(b) in Appendix). In particular, it can be seen that these one-gap models can only meet the specific heat data above 20 K and cannot satisfy the curves with negative curvatures or the bending down in low-temperature region. Considering the multiband feature of IBSs, the one-gap model is indeed difficult to account for this behavior. \par
	For a multiband system, electronic specific heat can be considered as a linear combination of multiple components. Each component has a gap structure and each gap has a partial Sommerfeld constant characterized as $\gamma_i$, thus $\sum\gamma_i = \gamma_n$. Then we use the two-gap model to fit the data. For the two samples, we fit it with (i) two isotropic $s$-wave gaps (see Fig.~\ref{fig7}(c) and 7(d) in Appendix), (ii) an isotropic $s$-wave gap plus an extended $s$-wave gap (see Fig.~\ref{fig4}(a) and 4(b)) and (iii) an isotropic $s$-wave gap plus a $d$-wave gap (see Fig.~\ref{fig4}(a) and 4(b)). We compare the three fitted results and come to a conclusion that the model with an isotropic $s$-wave gap plus an extended $s$-wave gap can fit the data best. For KCa$_2$Fe$_4$As$_4$F$_2$, the isotropic $s$-wave gap is 7.8 meV and the extended $s$-wave gap is 1.2 meV with $\alpha$ = 0.7. For KCa$_2$(Fe$_{0.95}$Ni$_{0.05}$)$_4$As$_4$F$_2$, the isotropic $s$-wave gap is 6.5 meV and the extended $s$-wave gap is 1.7 meV with $\alpha$ = 0.7. Both two samples have gaps with strong anisotropy and have deep gap minima of about 0.3-0.5 meV. We believe that the deep gap minima account for the behavior of $\Delta\gamma-H$ in Fig.~\ref{fig2}(c). To be precise, taking the parent sample as an example, up to 2 T, the field induced $\Delta\gamma(H)$ takes up 12.4$\%$ of the normal state specific heat coefficient $\gamma_n$, while it only increases to 16.7$\%$ $\gamma_n$ when the field rises up to 9 T.  This is because, a rather small magnetic field excites quasiparticles near the gap minima, so that the specific heat coefficient increases quickly with the magnetic field at first. Until a magnetic field up to about 2 T, the quasiparticles in the gap minima are almost excited sufficiently, and then $\Delta\gamma(H)$ steadily and slowly increases linearly with the magnetic field, returning to the normal behavior of an $s$-wave gap. After doping 5$\%$ Ni, the fitted $s$-wave gap becomes smaller, which is explained as the impurity scattering which decreases its $T_c$ value. While the fitted extended $s$-wave gap changes from 1.2 meV to 1.7 meV, which may be attributed to the electron doping of Ni modifying the topology of the Fermi surface and enlarging the electron pockets \cite{RN19}. We believe the smaller extended $s$-wave gap is from the electron pockets near $M$ point, and the larger isotropic $s$-wave gap comes from the hole pockets near $\Gamma$ point.\par
	Then we would like to discuss the fit of an isotropic $s$-wave gap plus a $d$-wave gap. Although this model can also roughly satisfy the data, it cannot be regarded as possible evidence that it has a $d$-wave component. Firstly, the total residual specific heat $\gamma_0$ plays an important role in determining the fitting results. In the fitting process, $\gamma_0$ is added to the BCS fitting curves to match the electronic specific heat data. This process results in a fitted $\gamma_0$. Additionally, in Fig. 2(a) and 2(b), $\gamma_0$ is determined using the formula $C(T,H) = \gamma_0(H)T + \alpha(H)T^2 + \beta T^3$. These different $\gamma_0$ values are illustrated in Table~\ref{tb2}. By comparing $\gamma_0$ presented in Fig.~\ref{fig2}(a) and 2(b), and $\gamma_0$ from BCS fitting, it becomes evident that the $\gamma_0$ value obtained from the $s + d$ fitting does not agree with the corresponding results from Fig.~\ref{fig2}(a) and 2(b). As a result, our analysis suggests that the $s$ + extended-$s$ fitting is more likely to be the prime candidate for the final results. From another perspective, ARPES \cite{RN14} and STM \cite{RN13} experiments suggest no existence of nodes, and thermal conductivity measurements \cite{RN11} and optical spectroscopy \cite{RN12} show the multiple nodeless $s$-wave gap in the 12442 system, which is consistent with our experimental results. $\mu$SR experiments \cite{RN15,RN16} successfully fit the experimental data with two $d$-wave gaps, but they study on the polycrystals. Impurities or the mixed orientations in polycrystalline samples may affect the results. In the Andreev reflection experiment for Rb12442 single crystal \cite{RN17}, the spectra display zero-bias peaks which suggest vertical nodal lines. That casts some doubt since in a point contact measurement, the zero-bias peaks can be induced by many reasons. In the specific heat measurement for K12442 single crystal \cite{RN18}, they speculate a possible $d$-wave by the existence of $\alpha(H)T^2$ term, but it could also be generated by some deep minima points or another relatively small gap. Thus, since there is no conclusive experimental proof of the nodal lines and $d$-wave pairing symmetry, or it is quite rare to report such nodal lines in IBSs, we would not believe the bending down in the low-temperature region is due to the existence of a $d$-wave gap, it most likely arises from the gap minima of an extended $s$-wave gap in the 12442 system.

\begin{table*}[htbp]
  \centering
  \caption{A summary of different $\gamma_0$ obtained from Fig.2(a) and 2(b) and obtained from different gap structures fitting.}
  \setlength{\tabcolsep}{3mm}{
    \begin{tabular}{c|c|l|r|c|l|r}
    \hline
    \hline
    \multirow{8}[6]{*}{$\gamma_0$(mJ/(mol K$^2$))} & \multicolumn{3}{c|}{KCa$_2$Fe$_4$As$_4$F$_2$} & \multicolumn{3}{c}{KCa$_2$(Fe$_{0.95}$Ni$_{0.05}$)$_4$As$_4$F$_2$} \\
  \hline
\cmidrule{2-7}          & \multicolumn{1}{l|}{From Fig. 2(a)} & \multicolumn{2}{c|}{From Fig. 4 and Fig. 7} & \multicolumn{1}{l|}{From Fig. 2(b)} & \multicolumn{2}{c}{From Fig. 4 and Fig. 7} \\
                         \hline
\cmidrule{2-7}          & \multirow{6}[2]{*}{24.3} & $s$     & 85.7     & \multirow{6}[2]{*}{43.9} & $s$     & 118.6 \\
          &       & $d$     & 61.0     &       & $d$     & 79.5 \\
          &       & extended-$s$   & 75.8     &       & extended-$s$   & 70.1 \\
          &       & $s + s$   & 51.2     &       & $s + s$   & 74.2 \\
          &       & $s + d$   & 17.4     &       & $s + d$   & 55.2 \\
          &       & $s$ + extended-$s$ & 22.6     &       & $s$ + extended-$s$ & 47.5 \\
    \hline
    \hline
    \end{tabular}}
   \label{tb2}
\end{table*}

\begin{figure}[htbp]
	\includegraphics[width=7cm]{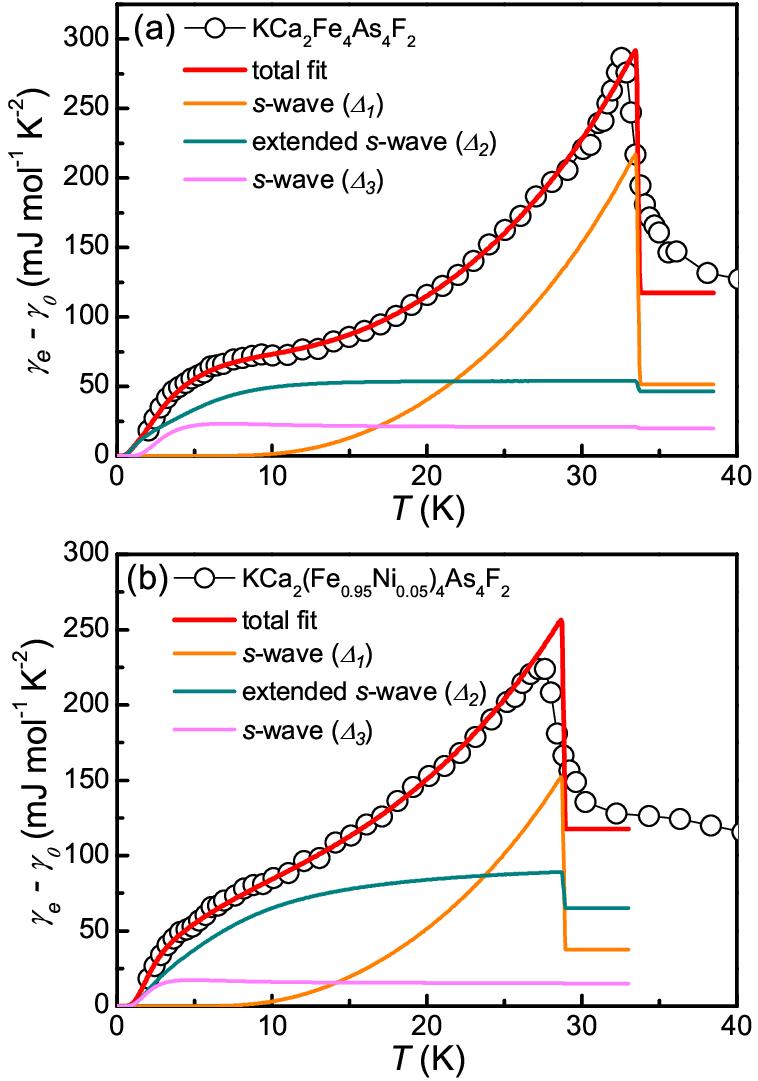}
	\caption{Temperature dependence of the electronic specific heat coefficient after subtracting the residual component under zero magnetic field for (a) KCa$_2$Fe$_4$As$_4$F$_2$ and (b) KCa$_2$(Fe$_{0.95}$Ni$_{0.05}$)$_4$As$_4$F$_2$ with a three-gap model. Symbols represent the experimental data. Solid lines represent fitting results and partial specific heat contributions of different gaps.} \label{fig5}
	\end{figure}

	Although the two-gap scenario has already satisfied the data, there remains a problem. We calculate the proportion of each gap and find that the large gap accounts for  43.9$\%$ and the small gap accounts for 56.1$\%$ in relative weight for KCa$_2$Fe$_4$As$_4$F$_2$, and the large gap accounts for 32.2$\%$ and the small gap accounts for 67.8$\%$ in relative weight for KCa$_2$(Fe$_{0.95}$Ni$_{0.05}$)$_4$As$_4$F$_2$. Since we attribute the contribution of the small gap to the electron pockets, the proportion seems too large in the 12442 system. There may be some small gaps existing in the hole-like pockets near $\Gamma$ point. \par
	According to ARPES experiment in KCa$_2$Fe$_4$As$_4$F$_2$ \cite{RN14}, they observe five hole-like pockets ($\alpha, \beta_1, \beta_2, \gamma_1, \gamma_2$) around $\Gamma$ point and small electron-like pockets ($\delta$) around $M$ point. The largest gap, approximately 8 meV, is found on the $\beta$ Fermi surface sheets, while the smallest gap, around 1 meV, is observed on the $\gamma$ sheets. It means that there exist large and small energy gaps in the hole pockets. In comparison with our specific heat data, the large gap of 7.8 meV in KCa$_2$Fe$_4$As$_4$F$_2$ is from the hole-like Fermi surface sheets, while the small gap with 56.1$\%$ weight should probably be the mixed contribution of electron and some hole pockets. Then we adopt a three-gap model to split the small gap into two different parts. In KCa$_2$Fe$_4$As$_4$F$_2$, we have reached the following results: three distinct gaps $\Delta_1$ = 7.8 meV, $\Delta_2$ = 1.5 meV, and $\Delta_3$ = 1 meV with 43.7$\%$, 39.5$\%$, and 16.8$\%$ in relative weight, respectively. We suggest that the isotropic $s$-wave gaps $\Delta_1$ and $\Delta_3$ are from the hole pockets at $\Gamma$ point, and the extended $s$-wave gap $\Delta_2$ with anisotropy $\alpha$ = 0.8 is from the electron pockets because of the complicated band structure around $M$ point. The large anisotropy, which causes a gap minimum of 0.3 meV, is necessary for the quasiparticle excitations, as we have mentioned above. The same fitting applies to KCa$_2$(Fe$_{0.95}$Ni$_{0.05}$)$_4$As$_4$F$_2$, resulting in $\Delta_1$ = 6.5 meV, $\Delta_2$ = 2 meV, and $\Delta_3$ = 0.7 meV. The magnitude of the largest gap $\Delta_3$ becomes smaller due to the decrease of $T_c$, and the magnitude of the extended $s$-wave gap $\Delta_2$ becomes larger due to the growth of the electron pockets after electron doping. The excellent fittings to the data for both samples are shown in Figs.~\ref{fig5}(a) and ~\ref{fig5}(b), which are plotted as $\gamma_e - \gamma_0$ vs $T$.\par
	In all, the fittings to the electronic specific heat again support the multigap features in the multiband 12442 system. We finally adopt a three-gap model, while a two-gap model can also fit the data well. The common point of the two treatments is that an extended $s$-wave gap with large anisotropy must be included, which we believe appears in the electron pockets.\par
	Based on the obtained fitting results, we observe the presence of non-zero residual specific heat coefficients $\gamma_0$ in our system. Next, we are going to discuss the possible origin of this residual specific heat coefficient.\par
	Firstly, one plausible explanation for $\gamma_0$ could be associated with a non-superconducting fraction in the superconductors. However, our high-quality single crystals demonstrate nearly full magnetic shielding volumes at low temperatures, suggesting minimal non-superconducting fractions, which effectively rules out this possibility. Consequently, a more viable explanation arises, suggesting that $\gamma_0$ originates from deep gap minima within a highly anisotropic extended $s$-wave gap.\par 
	In real materials, the existence of defects, disorders, and impurities is unavoidable. The impurity scattering effect can significantly influence the quasiparticle excitations in superconductors, particularly for superconductors with gaps that have nodes or deep minima points. Even in the low-temperature limit, the small gap amplitude near nodes or deep gap minima can be suppressed, resulting in a residual $\gamma_0$ value \cite{RN23}. On the contrary, in the case of an isotropic $s$-wave gap, the energy gap displays fully gapped characteristics around the Fermi surface, which does not allow to excite quasiparticles at low temperatures and typically does not contribute to a finite residual $\gamma_0$. In our case, the gap structure of K12442 superconductors is characterized by two $s$-wave gaps and one extended $s$-wave gap with deep gap minima. The large isotropic $s$-wave gap is unlikely to allow quasiparticle excitations at low temperatures. The small $s$-wave gap still has a size of about 1 meV and occupies a relatively small weight in the gap structure, thus it may not significantly contribute to the value of $\gamma_0$. Consequently, we attribute the majority of the $\gamma_0$ value to the highly anisotropic extended $s$-wave gap. 

\subsection{Condensation energy and its correlation with $T_c$}
\begin{figure}[htbp]
	\includegraphics[width=7cm]{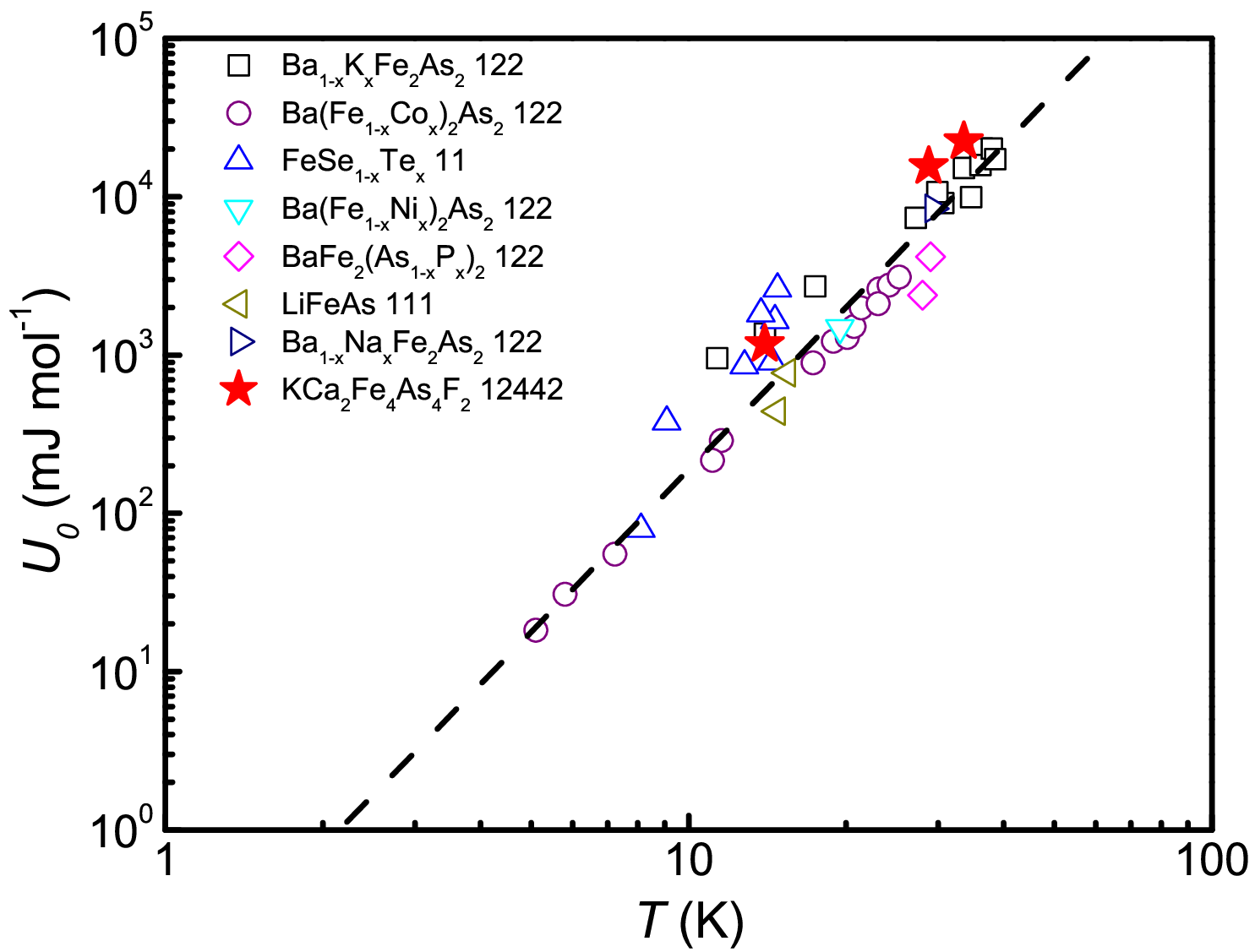}
	\caption{Correlations between the condensation energy and $T_c$ in some IBSs. Solid symbols represent data for KCa$_2$(Fe$_{1-x}$Ni$_x$)$_4$As$_4$F$_2$ with $x$ = 0, 0.5, 0.13 from our experiment, and the open ones are quoted from Ref. \cite{RN20} in which they summarize partial data from the available literatures \cite{RN20,RN26,RN27,RN28,RN29,RN30,RN31,RN32,RN33,RN34,RN35,RN36,RN37,RN38}. The dashed line indicates the relation $U_0 \propto {T_c}^{3.5}$.} \label{fig6}
	\end{figure}
	According to the thermodynamic formula, entropy is the partial derivative of Gibbs free energy with respect to temperature, and specific heat coefficient $C_{e}/T$ is the derivative of entropy with respect to temperature. Thus, we have the formula for condensation energy
     \begin{eqnarray}
U_0 = \int_{0}^{T_c}(S_n(T)-S_s(T))dT\\ \nonumber
 = \int_{0}^{T_c}dT \int_{0}^{T}\frac{C_n (T^{'})-C_s (T^{'})}{T^{'}}dT^{'}.  \label{equation6}
     \end{eqnarray}
	Therefore, we can obtain entropy by integrating the specific heat coefficient, and then obtain the condensation energy by integrating entropy to $T_c$. We calculate the condensation energy for KCa$_2$(Fe$_{1-x}$Ni$_x$)$_4$As$_4$F$_2$ with $x$ = 0, 0.5, 0.13. \par
	In order to understand the superconducting mechanism in present samples, we try to study the correlations between the condensation energy and $T_c$. In several IBSs, it is found that the specific heat jump exhibits a power law relation with the transition temperature $T_c$ observed by Bud'ko, Ni, and Canfield (BNC) \cite{BNC}, namely the BNC law ($\Delta C|_{T_c} \propto T_c^3$). In Ref. \cite{RN20}, the authors measured and collected a lot of data about the condensation energy and $T_c$ in several IBSs \cite{RN26,RN27,RN28,RN29,RN30,RN31,RN32,RN33,RN34,RN35,RN36,RN37,RN38}, and found an empirical relation of $U_0 \propto {T_c}^{3.5}$, which may have a close relationship with the BNC law. We also try to check this correlation with our data. As shown in Fig.~\ref{fig6}, our data fit the correlation quite well. It was mentioned in Ref. \cite{RN20} that, this unique correlation $U_0 \propto {T_c}^n$ ($n$ is about 3-4) is in clear contrast with the expectation of the BCS theory, in which it was predicted to be $U_0 = \frac{1}{2}N_F\Delta^2 \propto {T_c}^{2}$, thus $n$ = 2. The empirical relation was interpreted as a consequence of superconductivity occurring near the quantum critical point, thus the effective density of states changes with $T_c$ \cite{RN20}. Counting the hole concentration of our undoped samples, the doping level is about 0.25 hole/Fe for the undoped sample, which is close to the optimal doping level of about 0.2 hole/Fe. Doping Ni into the system usually means doping more electrons into the system, which ideally would lead to a doping level closer to the optimal point. However, the results show a decrease of $T_c$. Meanwhile, we see that the residual resistivity goes up quickly via Ni doping, showing an enhanced impurity scattering. As commonly accepted in the IBSs, the Ni ions in the system do not exhibit magnetic moments, thus the suppression to $T_c$ by doping Ni suggests that the gap should have a sign change. The consistency of our data with the empirical relation $U_0 \propto {T_c}^{3.5}$ strongly indicates that the correlation between condensation energy and $T_c$ reveals the intrinsic property in IBSs. Therefore, we believe that the specific heat of the 12442 system, as many systems in the iron-based superconducting family, is also beyond the understanding of the BCS theory and is intimately related to the unconventional superconducting mechanism.\par

\section{Conclusions}

	In summary, we investigate the specific heat of the multiband superconductors KCa$_2$(Fe$_{1-x}$Ni$_x$)$_4$As$_4$F$_2$ ($x$ = 0, 0.05, 0.13) single crystals and the overdoped non-superconducting single crystal with $x$ = 0.17. We observe the obvious specific heat anomalies at $T_c$ of 33.6 K and 28.8 K for the samples with $x$ = 0 and $x$ = 0.05, respectively. For the low-energy excitations, the magnetic field induced specific heat coefficient of the two superconducting samples increases quickly with the increasing field below 2 T, but shows a slow down and rough linear behavior above 2 T. Using the phonon contribution of specific heat of the non-superconducting sample as a reference, we successfully extract the electronic contributions of specific heat for the superconducting samples. By carefully identifying different fits of various pairing wave functions, we prefer the model with two $s$-wave gaps and an extended $s$-wave gap with large anisotropy. All the information above suggests that at least one anisotropic superconducting gap with deep gap minima exists in this multiband system. By Ni doping, $T_c$ decreases due to the impurity scattering effect, resulting in a decrease in the large $s$-wave gap, while the extended $s$-wave gap increases due to the enlargement of electron pockets. The general properties of the gap structure do not have a clear change with doping. Moreover, we calculate the condensation energy of the three superconducting samples, and find that it satisfies an empirical relation of $U_0 \propto {T_c}^n$ ($n$ = 3-4) proposed earlier, which is inconsistent with the traditional BCS framework.

\section*{ACKNOWLEDGMENTS}
We appreciate useful discussions with Gang Mu and Ilya Eremin. This work was supported by National Key R\&D Program of China (Grants No. 2022YFA1403200, and No. 2018YFA0704200), National Natural Science Foundation of China (Grants No. 11927508, 12061131001, No. 11974171, No. 11822411 and No. 11961160699), and the Strategic Priority Research Program (B) of Chinese Academy of Sciences (Grants No. XDB33000000). H. L. is grateful for the support from the Youth Innovation Promotion Association of CAS (Grant No. Y202001).

\section*{APPENDIX}
\begin{figure}[htbp]
	\includegraphics[width=8.5cm]{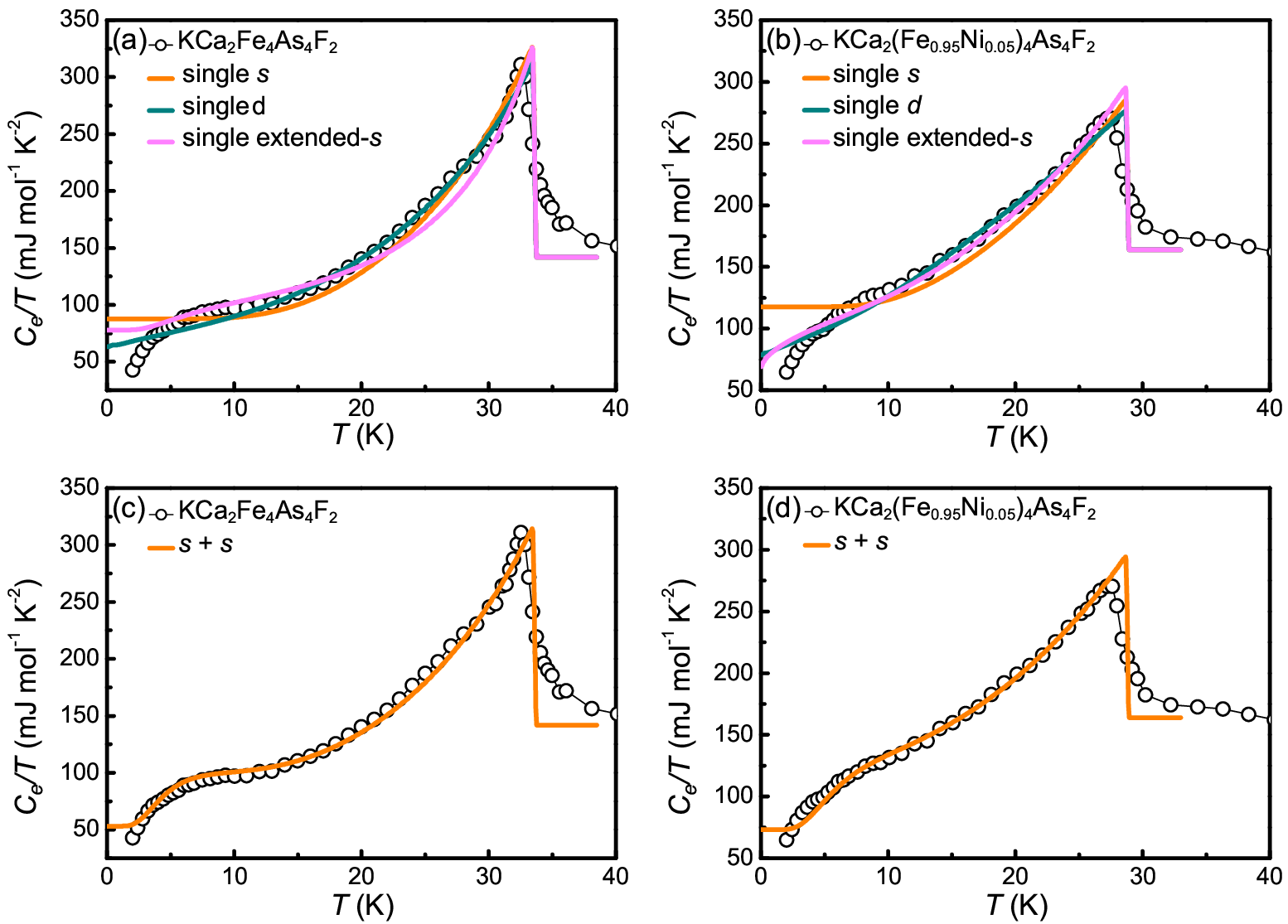}
	\caption{Temperature dependence of the electronic specific heat fitted with a one-gap model for (a) KCa$_2$Fe$_4$As$_4$F$_2$ and (b) KCa$_2$(Fe$_{0.95}$Ni$_{0.05}$)$_4$As$_4$F$_2$, with a two-gap model ($s + s$) for (c) KCa$_2$Fe$_4$As$_4$F$_2$ and (d) KCa$_2$(Fe$_{0.95}$Ni$_{0.05}$)$_4$As$_4$F$_2$. The lines represent the fitted curves with the BCS model.} \label{fig7}
	\end{figure}

\end{document}